\documentclass[twocolumn,secnumarabic,superscriptaddress,tightenlines,amssymb, aps, prb,showpacs]{revtex4b5}

\usepackage[oztex]{graphicx,color}

\begin{document}

\title{The Structure of Nanoscale Polaron Correlations in La$_{1.2}$Sr$_{1.8}$Mn$_{2}$O$_{7}$}

\author{B. J. Campbell}
\email{branton@anl.gov}
\affiliation{Materials Science Division, Argonne National Laboratory, Argonne, IL 60439}
\author{R. Osborn}
\affiliation{Materials Science Division, Argonne National Laboratory, Argonne, IL 60439}
\author{D. N. Argyriou}
\affiliation{Materials Science Division, Argonne National Laboratory, Argonne, IL 60439}
\author{L. Vasiliu-Doloc}
\affiliation{Department of Physics, Northern Illinois University, DeKalb, IL 60115}
\affiliation{Advanced Photon Source, Argonne National Laboratory, Argonne, IL 60439}
\author{J. F. Mitchell}
\affiliation{Materials Science Division, Argonne National Laboratory, Argonne, IL 60439}
\author{S. K. Sinha}
\affiliation{Advanced Photon Source, Argonne National Laboratory, Argonne, IL 60439}
\author{U. Ruett}
\affiliation{Materials Science Division, Argonne National Laboratory, Argonne, IL 60439}
\author{C. D. Ling}
\affiliation{Materials Science Division, Argonne National Laboratory, Argonne, IL 60439}
\author{Z. Islam}
\affiliation{Department of Physics, Northern Illinois University, DeKalb, IL 60115}
\affiliation{Advanced Photon Source, Argonne National Laboratory, Argonne, IL 60439}
\author{J. W. Lynn}
\affiliation{{NIST} Center for Neutron Research, National Institute of Standards and Technology, 
Gaithersburg, MD 20899}

\date{July 6, 2001}

\begin{abstract}
A system of strongly-interacting electron-lattice polarons can exhibit charge and orbital 
order at sufficiently high polaron concentrations.  In this study, the structure of short-range 
polaron correlations in the layered colossal magnetoresistive perovskite manganite, 
La$_{1.2}$Sr$_{1.8}$Mn$_{2}$O$_{7}$, has been determined by a crystallographic analysis of broad satellite maxima 
observed in diffuse X-ray and neutron scattering data.  The resulting $\mathbf{q}\approx (0.3,0,\pm 1)$
modulation is 
a longitudinal octahedral-stretch mode, consistent with an incommensurate Jahn-Teller-coupled 
charge-density-wave fluctuations, that implies an unusual orbital-stripe pattern parallel to the 
$<100>$ directions.
\end{abstract}


\maketitle
\section{Introduction}

The importance of electron-phonon coupling in amplifying the colossal magnetoresistive 
(CMR) effect in perovskite manganites was recognized at an early stage in the theoretical work 
of Millis and co-workers\cite{millis1,millis2}, who showed that the magnetic double-exchange mechanism\cite{zener,kubo}, which 
links the nearest-neighbor electron hopping rate to the degree of spin alignment, is not sufficient, 
by itself, to induce a metal-insulator transition.  In the insulating state above $T_{c}$, the strong 
coupling of the Mn$^{3+}$ $e_{g}$ electrons to Jahn-Teller distortions of the MnO$_{6}$ octahedra is essential in 
the formation of electron-lattice polarons, which are $e_{g}$ electrons that become trapped within their 
self-induced lattice distortions.\cite{roder}  The formation of the metallic state at $T_{c}$ marks the 
delocalization of the $e_{g}$ electrons and the collapse of their polaronic lattice distortions.\cite{tomioka,billinge,doloc}
If composition is expressed in terms of the Mn$^{4+}$ ion fraction, $x$, in these Mn$^{3+}$/Mn$^{4+}$ compounds (e.g. 
La$_{1-x}$Ca$_{x}$MnO$_{3}$ or La$_{2-2x}$Sr$_{1+2x}$Mn$_{2}$O$_{7}$), CMR behavior optimally occurs in the 
range $0.3 < x < 0.5$.

At high $e_{g}$-electron concentrations, where polarons interact via overlapping lattice-strain 
fields and electronic wave-functions, electronic and structural correlations can develop.  The case 
of long-range charge and orbital order (C/O), for example, may be viewed as one type of 
strongly-correlated polaronic limit.  While long-range polaron correlations are often absent at 
doping levels relevant to CMR effects, recent diffuse neutron and x-ray scattering experiments 
have revealed the existence of short-range polaron correlations that are intimately related to the 
behavior of the PI-FM transition.\cite{doloc,kubota,shimomura,adams,dai}
In La$_{1.2}$Sr$_{1.8}$Mn$_{2}$O$_{7}$, we have observed broad maxima in 
diffuse x-ray scattering data, approximately centered at reciprocal space positions $(h\pm 0.3,k,l\pm 1)$\cite{doloc}.  
These broad scattering maxima are diffuse surrogates of the satellite reflections that would be 
present if the underlying structural modulations produced by the polaron correlations were long-
range, and appear to be quasi-static on a 1 ps time scale.\cite{doloc}  Here we present a structural analysis 
based on the integrated intensities of a large number of these broad satellite maxima, providing a 
detailed description of the atomic displacements associated with the short-range polaron 
correlations above $T_{c}$.

\section{Experimental}

X-ray diffuse scattering measurements were performed using $6 \times 4 \times 1$ mm and 
$2 \times 2 \times 0.25$ mm single-crystal samples of La$_{1.2}$Sr$_{1.8}$Mn$_{2}$O$_{7}$ cleaved from the same region of a 
boule that was grown by the floating-zone technique.  Data were collected at 115 keV at the 
BESSRC 11ID beamline of the Advanced Photon Source using a Bruker 6500 CCD camera 
(sample to detector distance $= 2.5$ m), and at 36 keV at the SRI 1ID beamline using a Ge-solid-
state detector.

\begin{figure}[b]
\includegraphics{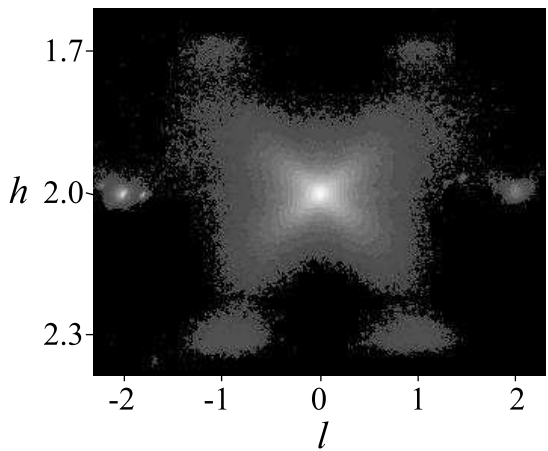}
\caption{Diffuse x-ray scattering data in the $k_{0}=0.005$ plane of reciprocal space simultaneously 
reveals polaron correlations, Huang scattering, and thermal diffuse scattering near the $(200)$ 
Bragg reflection.  The intensity scale of this CCD image is logarithmic, such that the strongest 
features are approximately 250 times more intense than the weakest features.}
\label{fig1}
\end{figure}

\section{Results and Discussion}

The diffuse scattering data in Figure \ref{fig1} were collected at 125 K using the CCD camera. Due to 
the small scattering angle $(2\theta\le 4.1^{\circ})$, this image represents, to a good approximation, a constant 
$k$ slice of reciprocal space centered at $(2,k_{0}=0.005,0)$.  The ``butterfly-shaped'' scattering pattern 
at the center of the figure is associated with the strain-fields induced by local JT distortions, and 
is commonly referred to as Huang scattering.\cite{huang}  The two narrow peaks at $(2,k_{0},\pm 2)$ are the tails of 
the $(2,0,\pm 2)$ Bragg peaks.  Most notably, the presence of the four diffuse maxima near 
$(2\pm 0.3,0,\pm 1)$ indicates the presence of the short-range polaron correlations.  Although the diffuse 
satellites are quite broad and also several orders-of-magnitude weaker in intensity than the parent 
Bragg peaks, they are sufficiently well-defined to reliably determine their integrated intensities. 
Because the peak widths did not vary significantly from satellite to satellite, a simple $h$-scan
through the center of each peak using the Ge solid-state detector was sufficient.  
The Huang scattering tails centered about nearby 
Bragg peaks produced a background that did vary significantly from one satellite to another.  
This background was accommodated by the peak fitting routine and subtracted.  The intensities 
of 109 unique diffuse satellites were thus measured at 125 K, and used to perform a 
crystallographic analysis with the JANA software package.\cite{jana}

The diffuse maxima observed at positions $\mathbf{Q}\approx \mathbf{Q}_{0}\pm(0.3,0,\pm 1)$, where 
$\mathbf{Q}_{0}$ is a Bragg peak 
position associated with the parent $I4/mmm$ symmetry (i.e. $h+k+l=2n$), can be equivalently 
described by positions $\mathbf{Q}_{0}+m\mathbf{q}$, where the modulation wave-vector is $\mathbf{q} \approx (0.3,0,0)$, and 
$\mathbf{Q}_{0}$ and $m$ 
are further restricted by the $(3+1)$-dimensional centering condition: $h+k+l+m=2n$. It is important 
to note that only first order satellite maxima (i.e. $m= \pm 1$) are observed.  Thus, they only appear 
adjacent to the $h+k+l = 2n+1$ positions.  This set of systematic absences leads to the selection of 
$Xmmm(\alpha 00)000$ as the $(3+1)$-dimensional superspace-group symmetry\cite{yamamoto,janssen}, where $X$ refers to the 
extended body-centering condition, $(x,y,z,t) \to (x+1/2, y+1/2, z+1/2, t+1/2)$, and implies that the 
modulated displacements in adjacent perovskite bilayers are 180$^{\circ}$ out of phase.  In the limit of 
small atomic displacements, an expression for the intensities of the diffuse satellite peaks can be 
written as:
\begin{equation}
\label{eq1}
I \propto |F|^{2}=\left| \frac{1}{2}(\mathbf{Q}_{0}+m\mathbf{q})\cdot \sum\limits_n \mathbf{u}_{n} f_{n} e^{i \mathbf{Q}_{0}\cdot
  \mathbf{x}_{n}}\right|^{2}
\end{equation}
where $\mathbf{x}_{n}$ and $f_{n}$ are the position and form factor of the $n$th atom in the unit cell, 
and $\mathbf{u}_{n} = \mathbf{u}_{c} + i\mathbf{u}_{s}$ is 
the modulation amplitude vector of the $n$th atom, which gives the displacement of the $n$th atom 
as $\mathbf{u}^{s}_{n} sin(\mathbf{q\cdot x}) + \mathbf{u}^{c}_{n} cos(\mathbf{q\cdot x})$.

\begin{figure}[t]
\includegraphics[height=15cm]{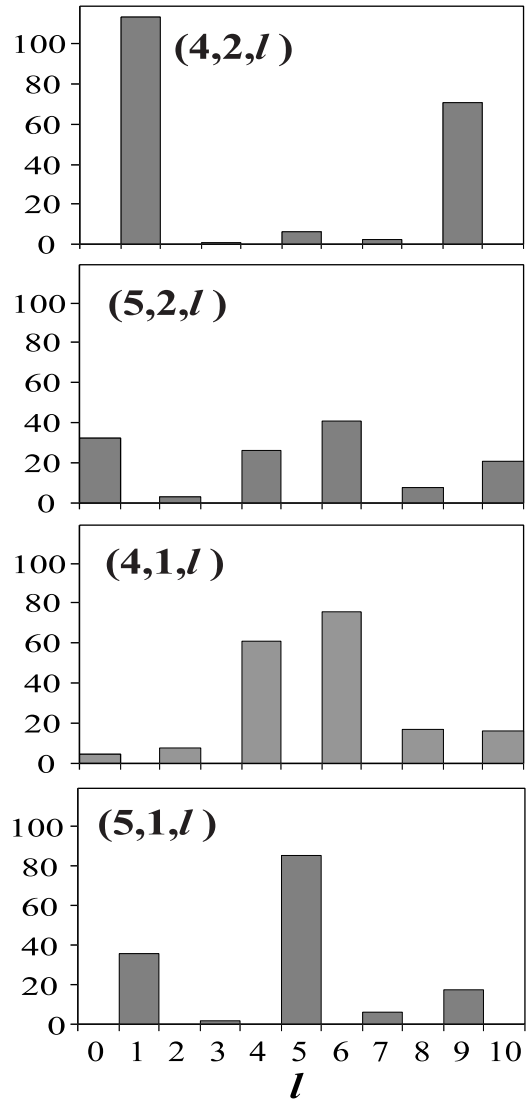}
\caption{A bar graph of $|F_{obs}|^2 \text{vs } l$ within each of the four distinct reflection subsets.  The $l$ 
dependence in each example is characteristic of its subset.}
\label{fig2}
\end{figure}

\begin{figure*}[t]
\includegraphics[width=\textwidth]{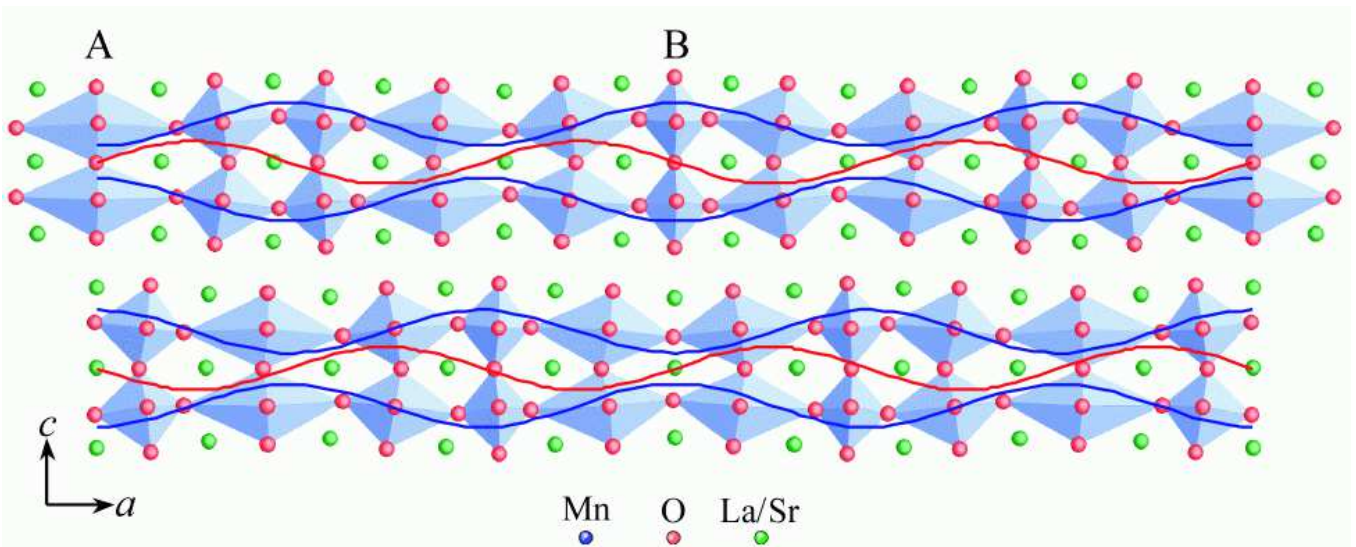}
\caption{Crystallographic representation of the one-dimensionally modulated structure associated 
with the diffuse $\mathbf{q}\approx (0.3,0,\pm 1)$ satellites. The atomic displacements are exaggerated in order to make 
the more subtle features of the modulation visually apparent.  The directions of the atomic 
displacements within each perovskite sheet and bilayer follow the blue ($z$-component) and red ($x$-
component) curves, where peaks indicate $+x$ or $+z$ displacements and troughs indicate $-x$ or $-z$ 
displacements.  The incommensurate modulation period has been approximated by the 
commensurate value of $\frac{10}{3}$.}
\label{fig3}
\end{figure*}

After the integrated intensities were extracted and corrected for absorption, several distinct 
features were immediately evident.  The intensities naturally separate into four distinct subsets, 
$(h,k) = {(e,o), (o,e), (e,e) \text{ and } (o,o)}$, where ``$o$'' and ``$e$'' indicate odd and even integers, 
respectively.\cite{oe}  For a given value of $l$, one subset is often notably more intense than the other. 
Because the O(3) oxygen is the only atom in the unit cell that produces a contribution to the 
structure factor capable of differentiating between these subsets, its displacement must be a key 
element of the modulation. The observed 1D modulation breaks the 4-fold symmetry of the 
average structure and splits the O(3) oxygen site in two distinct sites, referred to here as O(3a) 
and O(3b), where O(3a) connects two Mn atoms along the modulation direction, and O(3b) 
connects two Mn atoms along the direction transverse to the modulation.  The intensities within 
each of the four subsets increase strongly with increasing $h$, indicating that the principal 
displacements are parallel to the [100] modulation direction (i.e. longitudinal).  Furthermore, 
these intensities do not increase with $k$, indicating that there are no significant displacements 
along [010], which is also a requirement of the symmetry.

Distinctive intensity trends vs $l$ are observed within each of the four satellite subsets.  The 
trends are roughly periodic along $l$ with a period of roughly 10 reciprocal lattice units, a feature arising 
from the thickness $(\approx \frac{2c}{10})$ of the perovskite bilayers.  The details of this periodicity, and 
especially the differences amongst the four subsets, shed light on the structure of the modulation.  
Figure \ref{fig2} contains one characteristic example from each of the four satellite subsets.  
By examining these variations in light of the individual contributions to Eq. (\ref{eq1}) from each atom in 
the unit cell, we find that O(3a) experiences a large longitudinal displacement that is also in 
phase with those of its neighbors.  These observations serve as a starting model from which to 
refine all of the independent modulation amplitudes.

The structure in Figure \ref{fig3} illustrates the results of an $|F|^2$ refinement of the modulation 
amplitudes against the measured satellite intensities, which yielded an agreement factor of 
$(\sum w_{i}|\bigtriangleup I_{i}|^{2})/(\sum w_{i} I_{i}^{2}) = 16.3\%$.  
Only the $\mathbf{u}^{s}_{x}$ and $\mathbf{u}^{c}_{z}$ terms are permitted by symmetry\cite{petricek}, and both 
were essential to obtaining a good fit to the measured intensities.  The elongation of the 
Mn-O(3a) bonds at position A in the figure is interpreted as a cooperative Jahn-Teller (JT) 
distortion caused by the occupation of Mn $e_{g}$ orbitals with $d(3x^{2}-r^{2})$ character.  These Mn-O(3a) 
bond distortions are much more pronounced than any of the others.  Other cooperative features 
include the stacking of JT-distorted octahedra within a bilayer, the \textbf{\textit{c}}-axis compression of 
octahedra that experience \textbf{\textit{a}}-axis elongation, octahedral rotations about 
the \textbf{\textit{b}} axis, and the $180^\circ$ 
phase difference between the modulations in adjacent bilayers, which all appear to work together 
to minimize the lattice strain induced by the dominant Mn-O(3a) distortions.  After the 
sinusoidal modulation was applied, the model in Figure \ref{fig3} was stretched along [100] to restore the 
unphysically compressed octahedra at position B to their normal shapes.  This stretch 
corresponds to the local enlargement of the [100] cell parameter within the correlated regions 
due to the cooperative JT distortions.   Local lattice strain effects have been previously studied 
via pair distribution function analysis\cite{billinge}, whereas the present analysis is only 
sensitive to periodic features.

\begin{table}[t]
\caption{Average atomic coordinates\cite{xyz} and $\mathbf{q}\approx (0.3,0,\pm 1)$ modulation amplitudes.}
\label{tbl1}
\setlength{\tabcolsep}{6pt}
\begin{ruledtabular}
\begin{tabular}{llllll}
Atom & x & y & z & $\mathbf{u}^{s}_{x}$({\AA}$\times 10^2$) & $\mathbf{u}^{c}_{z}$({\AA}$\times 10^2$) \\
\hline
Mn & 0 & 0 & 0.0965 & 1.29( 3) & -1.03(12) \\
O(1) & 0 & 0 & 0 & 2.87(30) & \ \ \ \ \textemdash \\
O(2) & 0 & 0 & 0.1960 & 0.44(19) & -1.13(31) \\
O(3a) & 0.5 & 0 & 0.0952 & 4.69(20) & -1.73(35) \\
O(3b) & 0 & 0.5 & 0.0952 & 1.56(13) & -0.13(34) \\ 
La/Sr(1) & 0.5 & 0.5 & 0 & 1.31( 2) & \ \ \ \ \textemdash \\ 
La/Sr(2) & 0.5 & 0.5 & 0.1825 & 0.81( 2) & -1.54(45) \\
\end{tabular}
\end{ruledtabular}
\end{table}

The red ($x$-component) and blue ($z$-component) curves in Figure \ref{fig3} illustrate the relative phases 
of the modulation within each perovskite sheet and bilayer.  The curves are fixed by symmetry, 
and indicate that the \textbf{\textit{c}}-axis displacements within the two sheets of a perovskite 
bilayer are equal and opposite, whereas these two sheets share the same \textbf{\textit{c}}-axis 
displacements.  This structural nuance is qualitatively similar to one proposed by Kubota 
\textit{et al}.\cite{kubota}, based on single-crystal neutron 
diffuse scattering data from the related $x = 45\%$ system.  The refined values of the independent 
modulation amplitude components are listed in Table \ref{tbl1} in reciprocal lattice units, together with 
their respective atomic coordinates.\cite{xyz}  Since the amplitudes in Table \ref{tbl1} all have the same relative 
signs, a very important feature that is not dictated by symmetry, these curves also represent the 
cooperative displacement directions of all of the atoms in their respective layers at each point 
along the modulation.  Note that because the refinement scale factor is directly correlated to the 
modulation amplitudes in Eq. (\ref{eq1}), the values in Table \ref{tbl1} are defined only to within an overall scale 
factor, which was set by assuming a maximum Mn-O bond length distortion 
$(\bigtriangleup_{\text{Mn-O(3a)}})$ of 0.05\AA, a reasonable value based on the JT-distorted 
bond lengths observed for CE-type C/O order\cite{argyriou,kimura} in LaSr$_2$Mn$_2$O$_7$.  

\begin{figure}[b]
\includegraphics{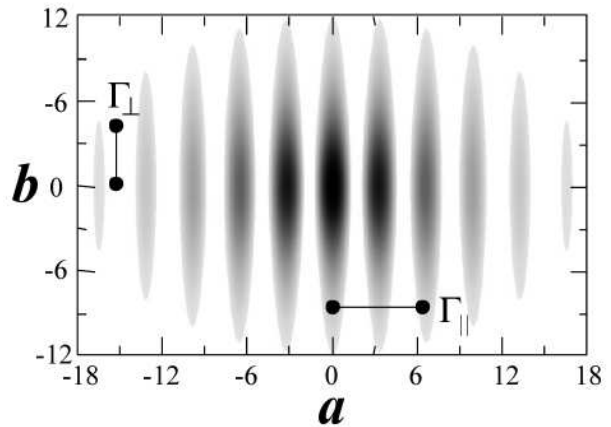}
\caption{The displacement-displacement correlation function associated with the short-range 
$\mathbf{q}\approx (0.3,0,\pm 1)$ modulation within a single perovskite plane, which takes into account the wave-
vector and the finite widths of the diffuse satellite peaks, and consists of a $cos^{2}(\mathbf{q\cdot r}/2)$ modulation 
factor multiplied by an exponential envelope, with HWHM values defined by the three 
independent correlation lengths $(\Gamma_{||}\approx 6a,\Gamma_{\perp}\approx 4a, \Gamma_{c}\approx c/2)$.  
Dark regions indicate high probability and light regions indicate low probability.  
The correlated regions are large enough to encompass several stripe like periods along [100].}
\label{fig4}
\end{figure}

The image in Figure \ref{fig4} represents the displacement-displacement correlation function associated 
with the short-range $\mathbf{q}\approx (0.3,0,\pm 1)$ modulation within a single perovskite sheet. 
Thus if the MnO$_{6}$ 
octahedron at the origin of the figure is Jahn-Teller distorted, the probability that another MnO$_{6}$ 
octahedron is similarly distorted will oscillate within the sheet along the modulation direction, 
and approach zero with increasing distance from the origin due to the finite range of the 
correlations.  When viewed in this fashion, the smoothly varying pattern of orbital stripes that 
forms perpendicular to the modulation direction is readily apparent.  However, one must note 
that the distribution in Figure \ref{fig4} is a purely statistical statement about the average size and shape of 
the correlated regions, rather than a picture of an individual correlation region.

The modulation illustrated in Figure \ref{fig3} challenges our current understanding of C/O order in 
CMR manganites.  First, it represents an orbital-stripe pattern parallel to the $<100>$ directions, 
and may be thought of as longitudinally-modulated ferroquadrupolar order.  This is in sharp 
contrast to the widely observed CE-type C/O order\cite{wollan,goodenough} and the related family of orbitally-striped 
phases\cite{chen,mori,radaelli}, in which the stripes lie parallel to $<110>$ directions and produce transverse, rather 
than longitudinal, displacements.  The CE-type C/O configurations support alternating rows of 
Mn$^{4+}$ and Mn$^{3+}$ sites as well as alternating $d(3x^{2}-r^{2})$ and $d(3y^{2}-r^{2})$ orbitals (i.e. charge + 
antiferroquadrupolar order), so as to minimize both Coulomb and lattice-strain energies.\cite{wollan,goodenough,mutou} 
Secondly, the modulation in Figure \ref{fig3} indicates a smoothly varying charge density.  Small Jahn-
Teller polaron models involve discrete Mn$^{3+}$ and Mn$^{4+}$ valence states, such that the Mn$^{3+}$O$_{6}$ 
octahedra are strongly distorted, while the Mn$^{4+}$O$_{6}$ octahedra retain their symmetry.  The 
incommensurate nature of the periodic Jahn-Teller distortions in Figure \ref{fig3}, however, indicates an $e_{g}$ 
electron density that varies continuously across the lattice, resulting in many MnO$_{6}$ octahedra 
with mixed Mn$^{3+}$/Mn$^{4+}$ characteristics.  And while the modulation satellites in La$_{1.2}$Sr$_{1.8}$Mn$_{2}$O$_{7}$ are 
quite broad, they are still much too narrow to be explained as an incoherent sum of scattering 
from distinct regions with commensurate modulation periods such as
3\textbf{\textit{a}} and 4\textbf{\textit{a}}.

A single-crystal neutron diffraction study of long-range CE-type C/O order at $x = 0.5$ 
concluded that the Jahn-Teller-distortions observed\cite{argyriou} in LaSr$_2$Mn$_2$O$_7$ were also too small to be 
explained in terms of discrete $3+$ and $4+$ valence states.  The smooth modulation of its charge 
and orbital degrees of freedom were instead interpreted as a weak charge-density wave (CDW) 
with $\mathbf{q}_{CE} = (1/4,1/4,0)$. Short-range CDW fluctuations are common in a variety of low-
dimensional systems, where they occur over an extended temperature range above a three-
dimensional ordering temperature, and give rise to diffuse reciprocal-space streaks similar to the 
diffuse peaks\cite{pouget} seen in Figure \ref{fig1}.  Their formation generally requires a peak in the electronic 
susceptibility at the CDW wavevector, usually a result of Fermi-surface nesting, together with 
strong electron-phonon coupling, which then permits a structural modulation to lift the nesting-
related degeneracy\cite{gruner}.  Furthermore, the $180^{\circ}$ inter-layer phase difference observed in Figure \ref{fig3} is a 
common feature\cite{gruner} of layered CDW systems that exhibit weak electron hopping between layers, 
which slightly corrugates the nested Fermi surfaces and thereby shifts the nesting condition by $q_{z} = \pm 1$.
In the case of La$_{1.2}$Sr$_{1.8}$Mn$_{2}$O$_{7}$, the modulation wave-vector does appear to be related to the 
electronic structure, as recent angle-resolved photoemission spectroscopy measurements and 
LSDA calculations reveal pronounced Fermi-surface nesting features in the metallic phase below 
$T_{c}$, with nesting vector $2\mathbf{q}_{F}\approx (0.3,0,\pm 1)$, while a wide pseudo-gap is observed to open above 
$T_{c}$.\cite{saitoh,chuang}  The long-range strain fields that produce the anisotropic butterfly scattering of Figure \ref{fig1} 
provide another important means of imposing nanoscale structure within the polaronic state.  
Because these strain fields overlap in $e_{g}$ electron-rich La$_{1.2}$Sr$_{1.8}$Mn$_{2}$O$_{7}$, a $\mathbf{q}$-dependence in the 
resulting strain-mediated inter-polaron interactions would also be expected to contribute to the 
structural modulation.

\section{Conclusion}

Rather than hosting independent polarons, an approximation that may be valid at low hole-
doping, the CMR manganites possess a dense population of polarons that interact via 
overlapping strain fields and electronic wave-functions, and might be described as polaronic 
liquids.  The novel $\mathbf{q}\approx (0.3,0,\pm 1)$ modulation uncovered in the present analysis is structurally 
consistent with a Jahn-Teller-coupled charge-density-wave fluctuation and possesses a plausible 
connection to Fermi-surface nesting features reported below $T_{c}$.  A longitudinal modulation of 
this nature has not been observed in the three-dimensional CMR manganites, and may be unique 
to lower-dimensional systems.  Yet, much remains to be discovered about the extent to which the 
$\mathbf{q}$-dependence of the strain-mediated inter-polaron interactions and the electronic susceptibility 
of the adjacent ferromagnetic metallic state play a role its development. The complex nanoscale 
structure observed within the polaronic state of La$_{1.2}$Sr$_{1.8}$Mn$_{2}$O$_{7}$ presents a challenge to our 
understanding and warrants further theoretical and experimental investigation into the nature of a 
concentrated population of strongly-interacting polarons.

\begin{acknowledgments}
This work was supported by U.S. DOE Office of Science under contract W-31-109ENG-38 
and by the State of Illinois under HECA.  We also acknowledge V{\'{a}}clav Pet{\u{r}\'{\i}\u{c}}ek (Institute of 
Physics AVCR, Czech Republic) for useful insights and assistance with the JANA software, 
James Phillips (Bruker-AXS, Madison WI) for assistance with the CCD camera, and J. C. Lang 
(SRI-CAT, Advanced Photon Source) for technical assistance at the 1ID beamline, as well as 
Richard Klemm, Michael Norman, and Dan Dessau  for helpful discussions.
\end{acknowledgments}

\bibliography{polaron}

\begin{thebibliography}{31}
\bibitem[{oe()}]{oe}
\bibinfo{note}{{Because} the diffuse satellites are located adjacent
  to the $h+k+l = 2n+1$ positions, $(h,k) = (o,e)$ not only indicates that $h$
  is odd and $k$ is even, but also implies that $l$ is even.}
\end{thebibliography}
\end{document}